# Comparative Density Functional Theory and Density Functional Tight Binding Study of Arginine and Arginine-Rich Cell-Penetrating Peptide TAT Adsorption on Anatase TiO$_2$

*Wenxuan Li, Konstantinos Kotsis, Sergei Manzhos[1]*

Department of Mechanical Engineering, National University of Singapore, Block EA #07-08, 9 Engineering Drive 1, Singapore 117576.

**Abstract**

We present a comparative Density Functional Theory (DFT) and Density Functional Tight Binding (DFTB) study of geometries and electronic structures of arginine (Arg), arginine adsorbed on the anatase (101) surface of titania in several adsorption configurations, and of an arginine-rich cell penetrating peptide TAT and its adsorption on the surface of TiO$_2$. Two DFTB parameterizations are considered, tiorg-0-1/mio-1-1 and matsci-0-3. While there is good agreement in the structures and relative energies of Arg and peptide conformers between DFT and DFTB, both adsorption geometries and energies are noticeably different for Arg adsorbed on TiO$_2$. The tiorg-0-1/mio-1-1 parameterization performs better than matsci-0-3. We relate this difference to the difference in electronic structures resulting from the two methods (DFT and DFTB) and specifically to the band alignment between the molecule and the oxide. We show that the band alignment of TAT and of TiO$_2$ modeled with DFTB is qualitatively correct but that with DFT using the PBE functional is not. This is specific to the modeling of large molecules where the HOMO is close to the conduction band of the oxide. We therefore report a case where the approximate DFT-based method - DFTB (with which the correct band structure can be *effectively* obtained) - performs better than the DFT itself with a functional approximation feasible for the modeling of *large* bio-inorganic interfaces, i.e. GGA (as opposed to hybrid functionals which are impractical at such a scale). Our results highlight the utility of the DFTB method for the modeling of bioinorganic interfaces not only from the CPU cost perspective but also from the accuracy point of view.

[1] Department of Mechanical Engineering, Faculty of Engineering, National University of Singapore, Block EA #07-08, 9 Engineering Drive 1, Singapore 117576. Tel: +65 6516 4605, fax: +65 6779 1459, E-mail: mpemanzh@nus.edu.sg .

## 1. Introduction

Bioinorganic materials and their interfaces have attracted significant attention, because they are promising materials in the areas of nanotechnology and biomedicine.[1-2] Applications of materials containing interfaces of metals or oxides with biomolecules range from dentistry[3] to drug delivery[4-5] to detection or sensing of molecules[6-12] and their conformers[13-14]. Studies of interactions of biomolecules with oxides and specifically titania are important for understanding biocompatibility of implants,[15] toxicity of environmental particles,[16] and for the development of essays and sensors.[17-18] Basic information about interaction of large biomolecules with metallic or oxide surfaces can be obtained by studying interactions with individual aminoacids, as is done in many studies.[19-30] While bioinorganic interfaces have been extensively studied experimentally,[13,15,21-23,25-30] experimental techniques do not provide a direct view of the molecular motions during adsorption, do not directly measure adsorption strength or geometry, and do not directly provide the mechanism of interaction. Simulations are therefore necessary to provide a detailed understanding of the interactions between the biomolecules and inorganic matter, which is necessary for rational rather than ad hoc design of new materials with desired functionality.

Because bioinorganic interfaces consist of an organic and an inorganic subsystems, their modeling poses challenges, as highly accurate methods have been developed for the modeling of individual components but not necessarily of their interfaces. For example, for molecular modeling, there exist a multitude of methods of different levels of accuracy and different CPU cost, starting from classical force field based molecular dynamics methods (FFMD)[31-35] to first-principles quantum chemical methods, such as density functional theory (DFT)[36-37] and various wavefunction based methods such as MP2[38-39] or the highly accurate coupled-cluster (CC)[40-42] methods. The wavefunction methods are of limited applicability to biomolecules as computational cost scales rapidly with the number of electrons $N_e$ (~$N_e^4$ and ~$N_e^6$ for MP2 and CCSD, respectively). DFT scales as ~$N_e^3$ and is feasible for rather large molecules with $10^{2-3}$ atoms such as peptides. Order-$N$ scaling DFT methods[43-45] have been developed but suffer from large prefactors. Due to their enormous CPU and memory cost advantage, force fields have been dominating the field of biomolecular simulation. The most commonly used types of force field such as CHARMM[46-51] or AMBER[52] represent key bonding and non-bonding interactions with simple parameterized functions (e.g. quadratic for covalent bonds or simple trigonometric functions for dihedral angles). Therefore, limited orders of coupling among nuclear degrees of freedom are included, and the simplicity of the functional form significantly limits the accuracy; patent failures of such FFs have been documented.[53]

For the modeling of inorganic solids and their surfaces, DFT as well as force fields are also used. However, both DFT and FF schemes which are optimal for solids are not the same as those for molecules. Specifically, as periodic boundary conditions (PBC) are typically applied to model bulk and surfaces, plane wave expansions of the orbitals and density are often used, with which it is necessary to use pseudopotentials (PP). This is often not necessary in molecular simulations. For the same reason, GGA (generalized gradient approximation) functionals such as the Perdew-Burke-Ernzerhof (PBE)[54] functional are typically used, while generally more accurate hybrid functionals are easier to apply to molecules. The net result is that it is difficult to use highly accurate DFT setups for bioinorganic interfaces. Even with a less costly setup such as a GGA functional and large core pseudopotentials, the computational cost of modeling such interfaces with DFT is large; specifically, slab surface models of lateral extent of about 30 Å are not *routinely* computable. However, such large slab sizes are required for an efficient modeling of bioinorganic interfaces in view of the large sizes of biomolecules.

Force fields[55] used for metals and semiconductors also differ significantly from those typically used for molecules and, specifically, biomolecules. This has to do in particular with the fact that while for molecular systems, interaction energies rapidly converge in an expansion over orders of coupling[56-59] of intramolecular coordinates, this is generally not the case in metals and semiconductors. Two-body terms are usually described by empirical pairwise interatomic parametrized potentials consisting of a long-range Coulomb potentials, a short-range Morse or Buckingham potentials, and Lennard-Jones potentials for the repulsive interactions.[60-64] Coupling terms are often described by density embedding schemes, such as the embedded-atom method (EAM) by Daw and Baskes and its variants.[65] As a result, it is difficult to model mixed systems with good accuracy also using force fields.

Overall, the accuracy of force field modeling can be much worse than that of DFT with errors in interaction energies and kinetic barriers on the order of tenth of an eV or even 1 eV.[66-67] Besides, FF simulations in principle cannot provide electronic structure and electron density information. It is an important limitation, as e.g. information about propensity to fold can be obtained by analyzing electron density.[68-69] Charges on atoms can be obtained from first principles if density is available. Further, interactions of biomolecules with inorganic matter can result in bond reformation which is difficult to account for with simple force fields. Reactive force fields[67,70] need to be used which are much more complex, often require tuning to a specific system, and may not approach ab initio accuracy.[67,71] Therefore, it is desirable to be able to perform electronic structure calculations on biomolecules and biomolecule containing interfaces, but such calculations are usually too costly at the DFT level.

DFTB (Density Functional Tight Binding)[72-80] is very attractive for the modeling of such interfaces because of a CPU cost which is about 3 orders of magnitude lower than that of DFT and amenability to near-linear scaling for large systems. The 3 orders of magnitude cost advantage is sufficient to routinely model sufficiently large slabs which can accommodate large biomolecules. DFTB therefore allows bringing ab initio accuracy and the ability to compute electronic properties to the modeling of systems of size which typically called for FF modeling. DFTB is an approximate DFT method which can provide high accuracy for system types for which it was parameterized, and it can be parametrized specifically for bioinorganic interfaces. There already exist DFTB parameterizations suitable for organic molecule-surface interactions, for instance DFTB was successfully employed to model acetic acid on $TiO_2$,[81] PTCDA (3,4,9,10-perylene tetracarboxylic dianhydride) on S-passivated GaAs,[82] and glycine on silica.[83] While as a method DFTB is certainly suitable for modeling of bioinorganic interfaces with DFT-like accuracy, a specific parameterization might not be.[84]

In this paper, we aim to assess the performance of DFTB with two parameter sets: matsci-0-3[85] and a combination of mio-1-1[72] and tiorg-0-1[86] when modeling arginine-rich cell-penetrating peptides. Arginine-rich cell-penetrating peptides are important for their capability to act as delivery vehicles,[87] which means they can deliver certain small molecular drugs and other therapeutic agents into the cytoplasm. Examples include the short arginine-rich transduction domain of the HIV-1 transactivator of transcription (TAT) protein.[88] It is worth noting that TAT has attracted the most attention as a prototypical example that has many of the essential characteristics of the arginine-rich cell-penetrating peptides.[89] Materials which contain titanium (Ti) are often found in biological environment, and their surfaces are usually oxidized under ambient conditions.[90] Thus understanding mechanistic details of the interactions between $TiO_2$ and Arginine is useful to understand the way peptides interact with oxides, which may help improve the design of safe nanomaterials that could find use in nanomedicine.[91] Specifically, we address the following questions: (i) what are adsorption energies and geometries of Arginine on anatase (101) surface of titiania (which is the majority surface)[92]? (ii) what are differences between DFT and DFTB models? (iii) how do pp. (i-ii) affect the adsorption of Arginine-rich peptides, i.e. what are the differences between the adsorption of a single aminoacid and a larger biomolecule? To help answer the last questions we also consider a dipeptide.

Because of the significant computational cost of DFT modeling of a peptide-titania interface, the strategy we adopt to answer these questions is (i) to study in detail the adsorption of arginine and arginine dipeptide on $TiO_2$ with both DFT and DFTB and to relate adsorption properties to the band structure and specifically band alignment of the molecule with titania, (ii) consider TAT adsorption on titania by modeling it directly with DFTB, and simulating indirectly the band alignment expected with DFT. The paper is organized as follows: In Section 2, we describe the DFT and DFTB methods and parameters used in the calculations; in Section 3, first the calculations of structures and band structures of arginine, arginine dipeptide, and the TAT peptide are presented, followed by calculations for the geometric and electronic structures as well as adsorption energies of different configurations of the arginine molecule and the dipeptide on the anatase (101) surface with the two methodologies (i.e. DFT and DFTB); finally, we compute the adsorption of TAT on $TiO_2$ with DFTB. Section 4 concludes.

## 2. Methods

DFT[36-37] calculations for isolated arginine, arginine dipeptide, and TAT molecules were performed with the Gaussian 09[93] (G09) program using the B3LYP[94-97] functional and the 6-31g+(d,p) basis set for the purpose of comparing the DFTB results with a quantitatively accurate DFT setup. For arginine, we have that the 6-31g+(d,p) basis set provides converged values vs. 6-31g++(2d,2p) (see Table 1).

DFT calculations of the adsorption of Arg and $Arg_2$ on $TiO_2$ were performed using the periodic slab surface model of titania. The SIESTA code[98] was used with the PBE[53] functional and a DZP (double-$\zeta$ polarized) basis. Troullier-Martins pseudopotentials[99] were used to account for core electrons. Basis functions of different width were used (as defined by specifying the PAO.EnergyShift parameter) to quantify possible artifacts due to the use of an atom-centered basis set. The tolerance for the relative change in the density matrix was set to $1\times10^{-5}$. The (101) surface of $TiO_2$ was modelled with a 8-Ti layer slab (144 atoms) of lateral size 11.3×10.3 Å and 26.0 Å of vacuum for arginine and 37.0 Å of vacuum for arginine dipeptide. The Brillouin zone was sampled with a 2×2×1 Monkhorst-Pack[100] grid point and a 200 Ry cutoff was used for the Fourier expansion of the electron density. The positions of atoms in the lower half of the slab were fixed at bulk positions, and all other atoms and lattice vectors were allowed to relax until forces were below 0.02 eV/Å and stresses below 0.1 GPa (for bulk $TiO_2$), respectively. The dipole correction[101-102] in the direction normal to the surfaces was checked but found to be insignificant.

SCC-DFTB[72-80] calculations were performed with the DFTB+ software[103]. We used two different parametrizations, matsci-0-3[85] and a combination of mio-1-1[72] and tiorg-0-1[86] (this combination will be referred to below as tiorg-0-1* for simplicity). A 8-Ti layer slab of lateral size 22.9×20.9 Å (576 atoms) and 13.4 Å of vacuum was used for arginine on $TiO_2$; the Brillouin zone was sampled with a 2×2×1 Monkhorst-Pack[100] grid of points. For TAT on $TiO_2$, a larger slab of 45.8×41.8 Å (2304 atoms) and vacuum layer of 60 Å were needed for the proper relaxation of the TAT molecule on the anatase surface, and only the tiorg-0-1* parametrization was used in the calculations because with these parameters we obtained more reliable results for arginine adsorption on anatase (101) surface. Here, the Brillouin zone was sampled at the Γ point due to the large size of the cell. Geometries were optimized until forces were below 0.02 eV/Å.

The adsorption energy is computed as:

$$E_{ads} = E_{titania/molecule} - E_{titania} - E_{molecule}$$

where $E_{titania/molecule}$ is the total energy of the bioinorganic interface system, $E_{titania}$ the total energy of the clean anatase (101) surface and $E_{molecule}$ the total energy of the isolated arginine, dipeptide or TAT molecule. A negative $E_{ads}$ value therefore corresponds to favored adsorption.

Previous ab initio studies imposed different charge states on the amino acids (e.g. on arginine molecule), including neutral.[19-20,104-112] Different codes used here allow for different treatments of solvent effects (PCM[113-117] in Gaussian, only explicit solvation in SIESTA and DFTB+, which is not practical for interfaces used here) which tend to produce different results beyond the differences induced by a specific DFT/DFTB approximation to the electronic structure. Therefore, to enable proper comparison between the two DFT setups and the DFTB setups, all systems (molecules and interfaces) were modelled in vacuum and in neutral state. For the same reasons, we do not include dispersion corrections. Even without such corrections, we obtain relatively strong chemisorption except for adsorption via amino groups in DFTB in a couple cases. We also focus our study on single molecule adsorption, i.e. the arginine molecule and the dipeptide arginine as well as the larger rich-arginine peptide molecule in form of TAT. Only the linear conformer of Arg was considered, similarly to all previous works studying arginine adsorption[20,107-110,112] on $TiO_2$, and because it is the conformation it assumes in TAT (see Figure 2).

Trial TAT structures were generated using PEP-FOLD[118-119] and optimized with DFTB, whereby the lowest energy conformer was chosen and used in all calculation.

## 3. Results

### 3.1. Molecular calculations.

We first compare molecular and electronic structures of molecules: the Arginine aminoacid, Arginine dipeptide, and TAT, as obtained with DFT and DFTB. In Figure 1, we show the structures of these molecules optimized with DFTB (using the mio-1-1 parameter, set which is matched with tiorg-0-1 used for $TiO_2$ calculations, as well as the matsci-0-3 parameter set) and with DFT using two setups: the higher-accuracy Gaussian setup with the hybrid B3LYP functional suitable for molecules, as well as the SIESTA with the PBE functional suitable for studying adsorption on $TiO_2$. The wireframe structures in Figure 1 (a-c) are computed with, respectively, DFT/SIESTA, DFTB/mio-1-1, and DFTB/mastic-0-3 and are overlaid with structures computed with the more accurate Gaussian setup. The agreement is good with noticeable differences observed only in angular coordinates. The differences in bond lengths and angles for Arginine between these computational schemes are given in Table 1 together with differences in structural parameters related to the bond between two aminoacids in the dipeptide (Figure 2). These comparisons are expected to hold as well for TAT (also shown in Figure 2) and other peptides. From Table 1 and Figure 1 one can conclude that all four computational schemes provide largely similar structural information, in particular, that DFTB provides similar accuracy to DFT and especially mio-1-1 parameter set can provide more similar results to DFT compared with matsci-0-3. The largest differences are observed in dihedral angles and can reach 29 degrees with DFTB (matsci-0-3). Table 1 also shows differences in energy between two conformers of Arg (shown in Figure 1(a-c) and 1(d), respectively). The differences observed in Table 1 between the most accurate computational setup we used here (i.e. Gaussian) and which is typical of those used for the modeling of small molecules, and the DFT setup (SIESTA) typical of those used for the modeling of bulk materials and interfaces is an indication of what kind of accuracy can be expected from DFT when modeling bio-inorganic interfaces; this should be kept in mind when comparing DFT and DFTB results for adsorbate systems in the next section. Table 1 also shows that our G09 and SIESTA results are converged with respect to the choice of the basis.

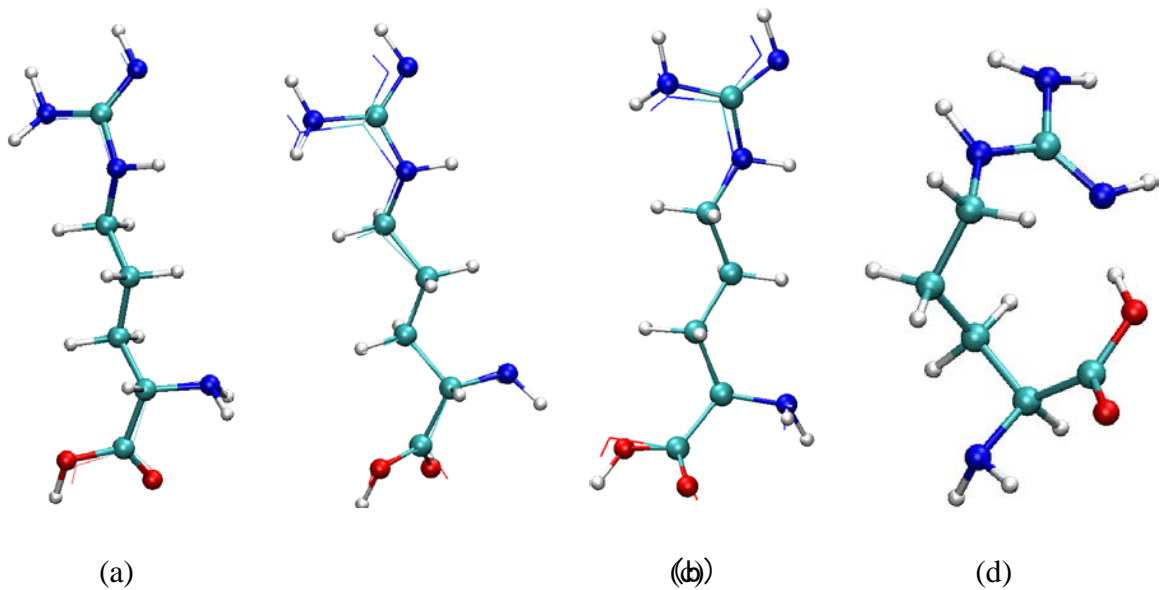

(a)  (b)  (c)  (d)

**Figure 1.** Wireframe optimized structures of Arginine with (a) DFT/SIESTA, (b) DFTB/mio-1-1, and (c) DFTB/matsci-0-3, overlaid with ball and stick optimized structures from DFT/Gaussian. Panel (d) shows the lowest-energy structure from Ref. [104]. The atom color code here and elswehere: C, green; O, red; H, light grey; N, blue.; Ti, dark grey. Visualization here and elsewhere by VMD[120].

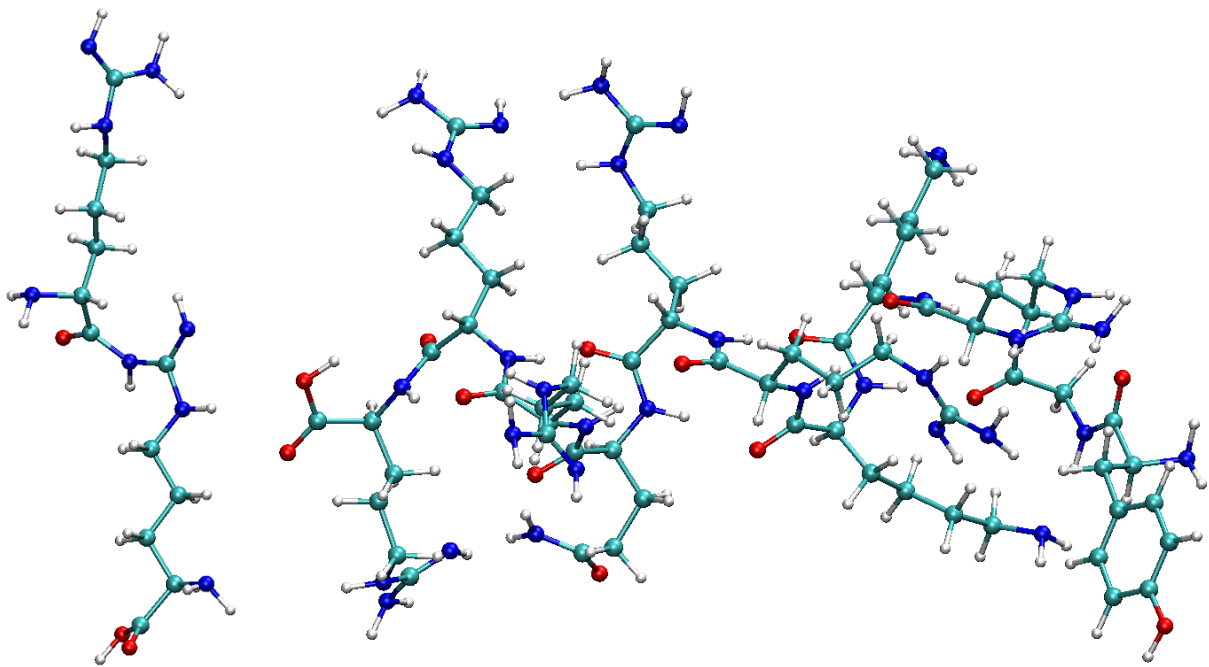

**Figure 2**. Structures of Arginine dipeptide (left) and the TAT peptide (right). The structures computed with Gaussian 09 are shown and are visually similar for all computational setups used here.

**Table 1.** Mean Absolute Difference of bond lengths (Å) and angles (°) for Arginine ($MAD_{bond}$ and $MAD_{angle}$) and the length of the peptide bond ($\Delta B_{pep-bond}$) between the Gaussian setup and the other three computational schemes, as well as the energy difference between the conformer shown in Figure 1 (a-c) and the cyclic structure shown in Fig. 1(d) ($\Delta E_{conformer}$, in eV).

|                       | $MAD_{bond}$ | $MAD_{angle}$ | $\Delta B_{pep-bond}$ | $\Delta E_{conformer}$ |
|-----------------------|--------------|---------------|-----------------------|------------------------|
| DFT/G09/6-31g+(d,p)   | /            | /             | /                     | 0.07                   |
| DFT/G09/6-31g++(2d,2p)| 0.002        | 0.08          | -0.004                | 0.07                   |
| DFT/SIESTA[a]         | 0.011        | 0.44          | 0.008                 | 0.34                   |
| DFT/SIESTA[b]         | 0.011        | 0.43          | 0.011                 | 0.35                   |
| DFTB/Mio-1-1          | 0.013        | 1.30          | 0.006                 | 0.20                   |
| DFTB/Matsci-0-3       | 0.033        | 1.77          | -0.028                | 0.15                   |

[a] Broader basis functions (PAO.EnergyShuft = 0.001 Ry)

[b] Narrower basis functions (PAO.EnergyShuft = 0.002 Ry)

In Figure 3, we compare the densities of states of the Arg, Arg dipeptide, and TAT following from the four methods. We see that the HOMO-LUMO gap narrows with the increase of the size of the molecule due to both an increase of the HOMO energy and decrease of the LUMO energy. Specifically, the HOMO of TAT is higher than that of Arg by 0.66, 0.85, 0.82, and 0.90 eV with DFTB/matsci-0-3, DFTB/mio-1-1, DFT/G09, and DFT/SIESTA, respectively, while the LUMO is lowered by 1.13, 1.32, 0.54, and 0.81 eV, respectively. What we will show is consequential to the modeling of peptide adsorption on oxides is the fact that the gap of a large biomolecule, TAT in this case, is much smaller than that of individual aminoacids. We will show in what follows that this has a significant effect on band alignment with titania and also affects the performance and applicability of DFTB *and DFT* setups. The band gap obtained with SIESTA is much smaller than with Gaussian or DFTB due to the use of the PBE functional, as expected[121]. The dotted line on the "Gaussian" panel of Figure 3 illustrates this by showing the DOS obtained for Arg in Gaussian with the PBE functional.

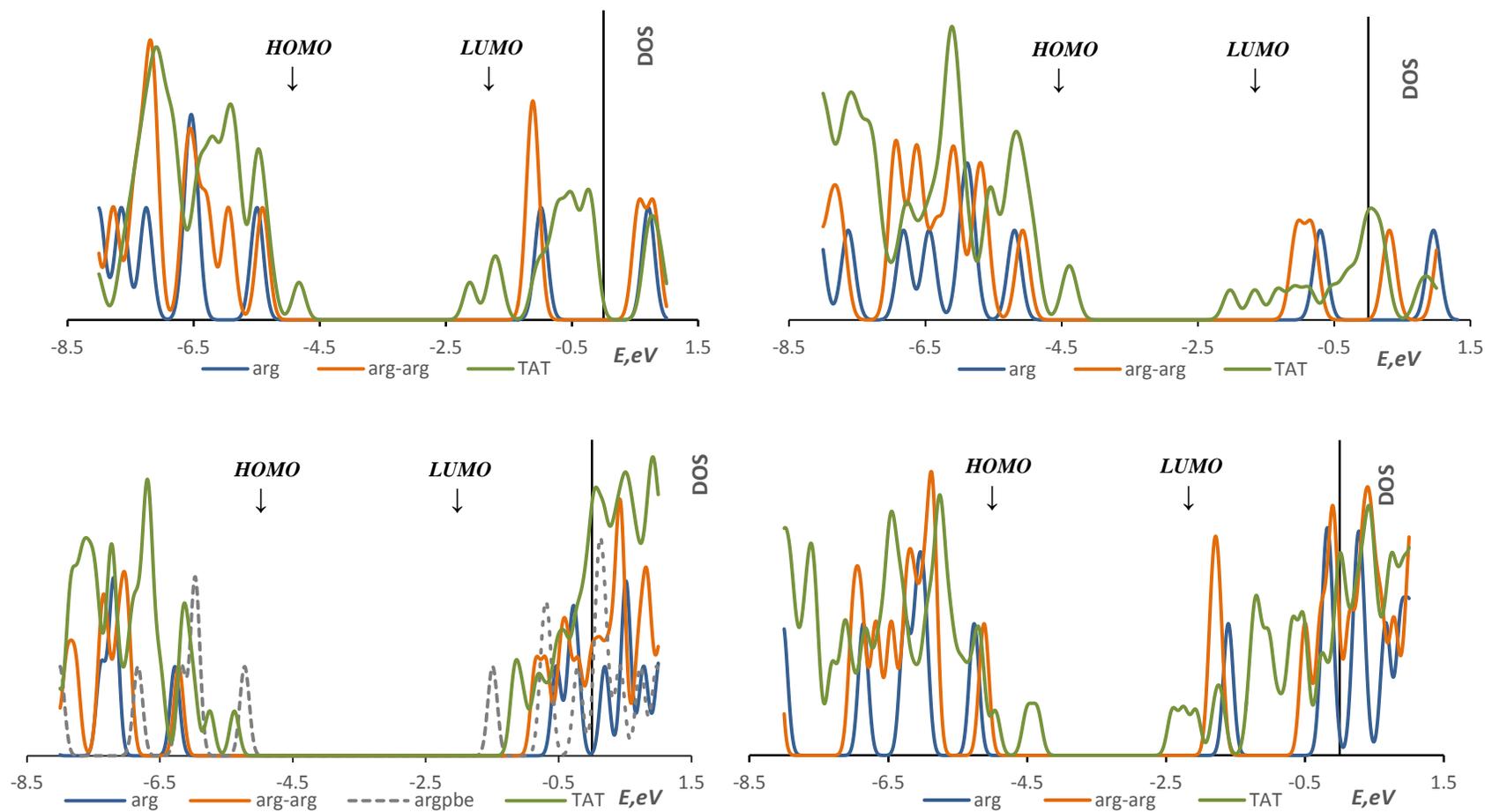

**Figure 3**. Densities of states of Arg, Arg dipeptide and TAT computed with DFTB (top left: matsci-0-3 and top right: mio-1-1 parameter sets) and DFT (bottom left: G09/B3LYP, bottom right: SIESTA/PBE). A Gaussian broadening of 0.1 eV is applied. Approximate positions of HOMO and LUMO energies are indicated with arrows.

### 3.2. Arginine-titania interfaces

We considered two adsorption regimes: adsorption via the carboxylic group and via the amine groups. The adsorption configurations via the –COOH groups obtained with DFTB and DFT (SIESTA) are shown in Figure 4. The corresponding adsorption energies ($E_{ads}$) and key bond lengths between the molecule and $TiO_2$ are given in Table 2. Similar to the molecular calculations, we have checked that adsorption geometries and energies obtained in SIESTA are sufficiently stable with respect to the choice of basis parameters, see Table 2 where results with two bases are given.

Three types of adsorption configurations are considered: a bidentate (BB) and two monodentate (M1 and M2) configurations, shown in Figure 4 (a, d) and Figure 4 (b, c, e, f), respectively. The DFTB structures in Figure 4(a-c) are obtained with the tiorg-0-1* parameterization and are visually similar to those with matsci-0-3 except that with matsci-0-3, in monodentate configurations, the hydrogen atom of the carboxylic group remains on the molecule (as in Figure 4(e-f)) while it dissociates with tiorg-0-1*.

In Table 2, significant differences in adsorption geometries and energies are observed among the three computational schemes. Specifically, the difference in the carboxylic group's hydrogen coordination between tiorg-0-1* and matsci-0-3 (DFTB) is reflected in different distances labeled as H-$O_{surf}$ in Table 2. Here, the H-$O_{surf}$ distances obtained in DFTB with matsci-0-3 are closer to the DFT results than those with tiorg-0-1* because with matsci-0-3 the H atoms remained on the molecule, similar to DFT. However, in previous ab initio works on molecules adsorbed on anatase (101) via the COOH group, H was or was not dissociated depending on the scheme employed.[122-123]

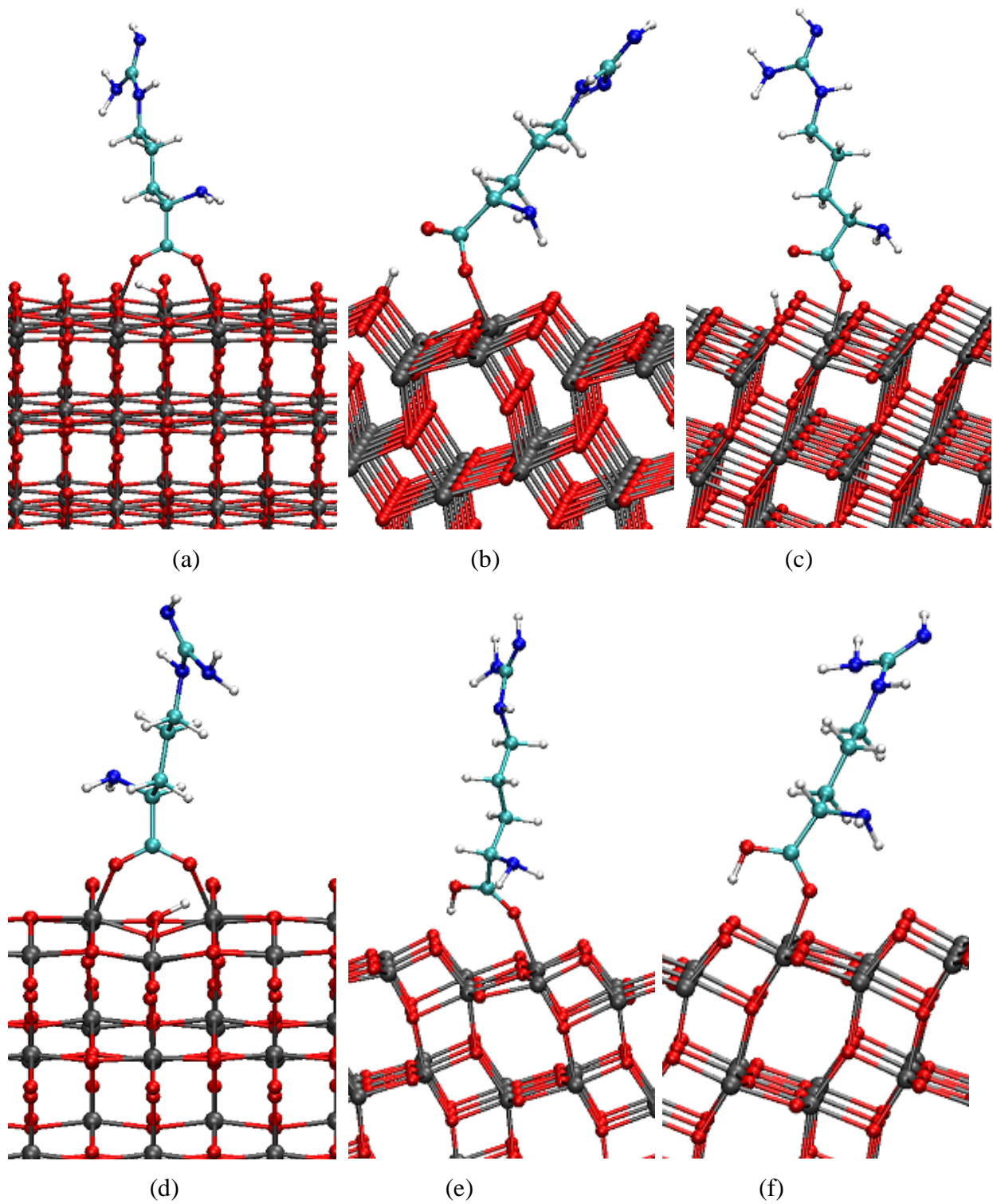

**Figure 4.** Optimized structures of BB (a), M1 (b) and M2 (c) configurations of Arg on TiO$_2$ from DFTB tiorg-0-1* and BB (d), M1 (e) and M2 (f) from SIESTA.

The $O_{mol}$-Ti bond lengths, which determine the electronic coupling between the molecule and the surface, with tiorg-0-1* are closer to the DFT results than those with matsci-0-3. The $O_{mol}$-Ti bond lengths with tiorg-0-1* straddle those obtained with DFT; specifically, in the BB configuration, they are slightly higher, and in the monodentate configurations, lower.

The adsorption energies obtained with tiorg-0-1* are closer to the DFT results than those obtained with matsci-0-3: $E_{ads}$ is about -1.0, -1.0, and -1.1 eV in BB, M1, and M2 configurations, respectively, vs. the DFT values of -1.0, -1.1, and -1.2 eV, respectively. Specifically, the order of adsorption strength among different configurations is reproduced reasonably well. In contrast, matsci-0-3 results in the strongest adsorption in BB ($E_{ads}$=-1.8 eV), followed by M1 (-1.0 eV) and M2 (-1.1 eV). The differences are therefore qualitative. Overall, for a single aminoacid, tiorg-0-1* provides more similar adsorption energies and geometries to DFT and will therefore be used below in calculations involving the dipeptide and the TAT peptide.

Figure 5 shows the molecule- and titania- projected density of states obtained for different adsorption configurations with the three computational schemes. Both DFTB with tiorg-0-1* and DFT have molecular HOMO in the bandgap and LUMO in the conduction band, except that in the BB configuration with tiorg-0-1*, the molecular HOMO remains just below the valence band maximum (VBM). In contrast, the molecular HOMO is inside the VB with matsci-0-3 for all adsorption configurations. All three approaches put the molecular LUMO in the conduction band of titania.

From Table 2, we also see that the monodentate configurations obtained with tiorg-0-1* (i.e. those for which the molecular HOMO is in the gap, similar to the DFT results) have smaller $O_{mol}$-Ti bond lengths than DFT-optimized systems, while the BB configuration (i.e. that for which the molecular HOMO is in the VB) – larger, similar to the overestimation of the $O_{mol}$-Ti bond lengths with matsci-0-3 (with which the molecular HOMO is in the VB in all configurations). We therefore see that the match between DFT and DFTB results depends on the qualitative match of their respective band structures, specifically, the band alignment between the molecule and the substrate. A similar observation has been made in other comparative DFT-DFTB studies of molecule-titania interfaces[81,84]. We note that the DFT model puts the molecular

HOMO only 0.62, 0.69, and 0.75 eV below the CBM for the BB, M1, and M2 configurations, respectively.

**Table 2.** Adsorption energies $E_{ads}$ (in eV) of Arginine in different configurations on anatase (101) surface of $TiO_2$ for adsorption via the carboxylic group. The bond length for bonding between the molecule's and surface atoms are also given (in Å). For bidentate configurations, the two bond lengths are $O_{mol}$-Ti; for monodentate, they are $O_{mol}$-Ti and H-$O_{surf}$. The data are for DFTB calculations with two parameterizations (matsci-0-3 and tiorg-0-1*) and for DFT calculations with different choices of DZP basis parameters (using PAO.EnergyShift of 0.001 and 0.002 Ry[98])

| System | $E_{ads}$, eV | $O_{mol}$-Ti, Å | $O_{mol}$-Ti/H-$O_{surf}$, Å |
|---|---|---|---|
| DFTB/Matsci-0-3 | | | |
| BB | -1.78 | 2.23 | 2.23 |
| M1 | -0.97 | 2.25 | 1.63 |
| M2 | -1.08 | 2.26 | 1.62 |
| DFTB/Tiorg-0-1* | | | |
| BB | -0.97 | 2.10 | 2.11 |
| M1 | -0.99 | 1.92 | 1.00 |
| M2 | -1.07 | 1.98 | 0.99 |
| SIESTA (0.001Ry) | | | |
| BB | -1.02 | 2.04 | 2.06 |
| M1 | -1.14 | 2.14 | 1.51 |
| M2 | -1.21 | 2.12 | 1.53 |
| SIESTA (0.002Ry) | | | |
| BB | -1.00 | 2.05 | 2.07 |
| M1 | -1.09 | 2.14 | 1.50 |
| M2 | -1.17 | 2.13 | 1.55 |

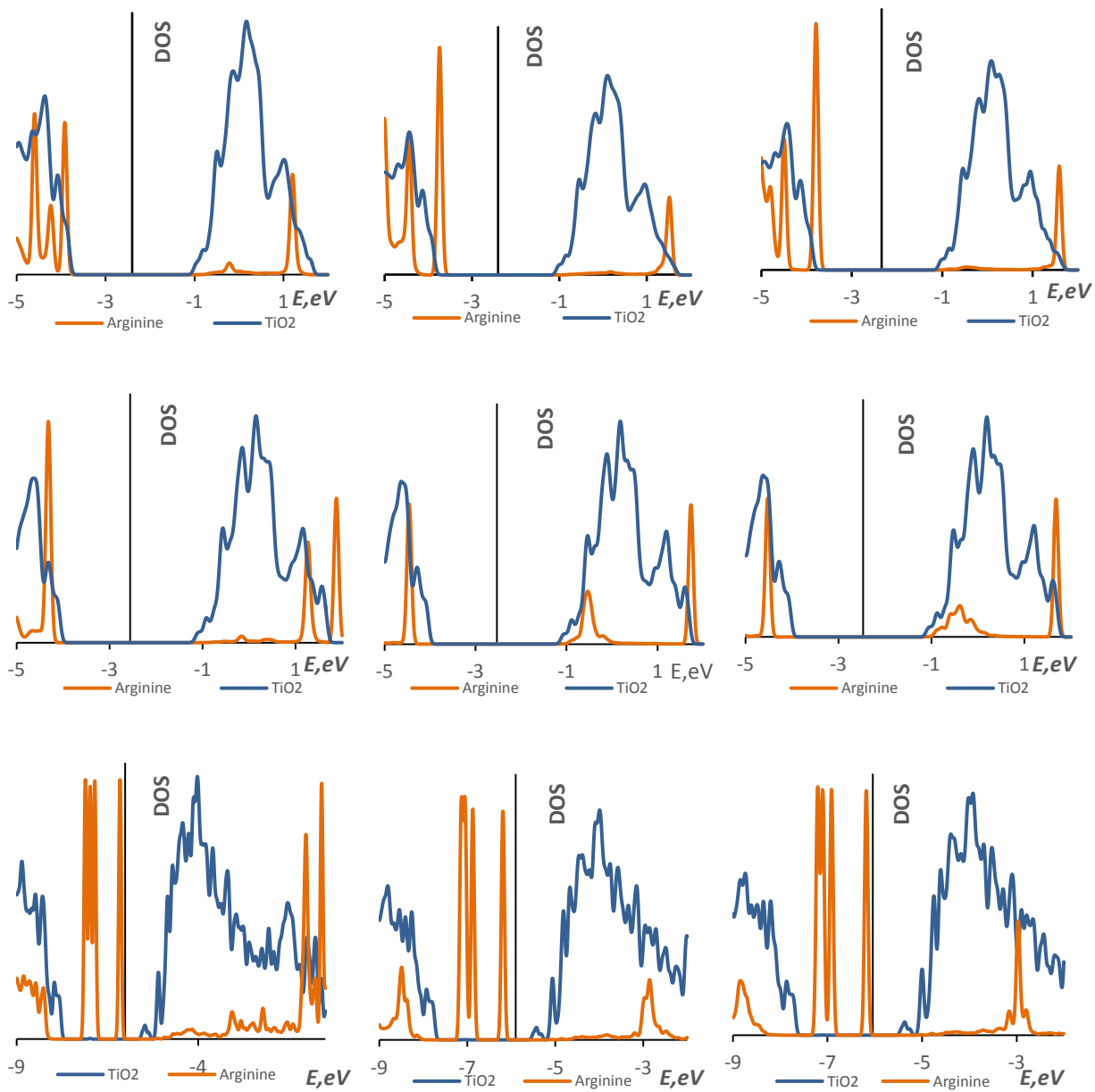

**Figure 5.** Molecule- and substrate- projected density of states of (left to rigth) BB, M1 and M2 obtained in DFTB with the tiorg-0-1* parameters (top row), matsci-0-3 (middle row), and with DFT (bottom row). The *y* axis crosses the energy axis at the Fermi energy.

Table 3 lists adsorption geometries and energies obtained with the three computational schemes for adsorption via the amine group, for which the geometries are shown in Figure 6. The adsorption can result in molecule's H atoms coordinating to surface oxygen atoms (these configurations are labeled as HO) or in molecule's N atom binding to Ti atom (these configurations are labeled as NTi). In Figure 7 shown are titania- and molecule- projected densities of states obtained with the three approaches. Similar to the adsorption via the carboxylic group, DFTB calculations with the tiorg-0-1* parameter set put the molecular HOMO into the bandgap, while those with matsci-0-3 put the HOMO inside the valence band. The band alignment obtained with tiorg-0-1* matches therefore qualitatively that obtained with DFT. The order of adsorption energies between HO and NTi configurations obtained in DFTB with tiorg-0-1* also matches that obtained with DFT although the magnitude of $E_{ads}$ is much weaker. The order of $E_{ads}$ with matsci-0-3 does not match that of DFT. The conformation assumed by the molecule in the OH configuration with matsci-0-3 is also very different to that obtained with DFT and with tiorg-0-1* (even though both tiorg-0-1* and matsci-0-3 calculations used the same DFT-based initial configuration). Similarly to the adsorption via COOH, there is therefore correspondence between ability to qualitatively reproduce the band alignment and adsorption properties. Similarly to the adsorption via COOH, DFT puts the HOMO very close to the CBM of titania, only 0.23 eV below it, in the HO configuration, while the DOS observed with the NTi configuration shows the HOMO near VBM. The PDOS of the NTi configuration is therefore different (quantitatively) to those of other adsorption configurations. This is because of the geometry the molecule assumes in this configuration. This was confirmed by the comparison of the DOS of a single arginine molecule in vacuum in its equilibrium configuration and in configurations it assumes during adsorption.

**Table 3.** Adsorption energies $E_{ads}$ (in eV) of Arginie in different configurations on anatase (101) surface of TiO$_2$ via the amine groups. The bond length for bonding between the molecule's and surface atoms are also given (in Å). For H-O configurations, the two bond lengths are O$_{surf}$-H; for NTi configurations, they are N$_{mol}$-Ti. (The bond length cannot be defined in the same way for the HO configuration obtained with matsci-0-3 due to a very different resulting geometry.)

| System | $E_{ads}$, eV | O$_{surf}$-H/N$_{mol}$-Ti, Å | O$_{surf}$-H/N$_{mol}$-Ti, Å |
|---|---|---|---|
| DFTB/Tiorg-0-1* | | | |
| H-O | 0.04 | 1.97 | 2.08 |
| N-Ti | -0.22 | 3.49 | 3.78 |
| DFTB/Matsci-0-3 | | | |
| H-O | -2.16 | / | / |
| N-Ti | -1.27 | 2.27 | 5.11 |
| SIESTA/0.001Ry | | | |
| H-O | -0.26 | 2.14 | 2.11 |
| N-Ti | -1.55 | 2.12 | 3.89 |
| SIESTA/0.002Ry | | | |
| H-O | -0.21 | 2.14 | 2.09 |
| N-Ti | -1.65 | 2.08 | 4.08 |

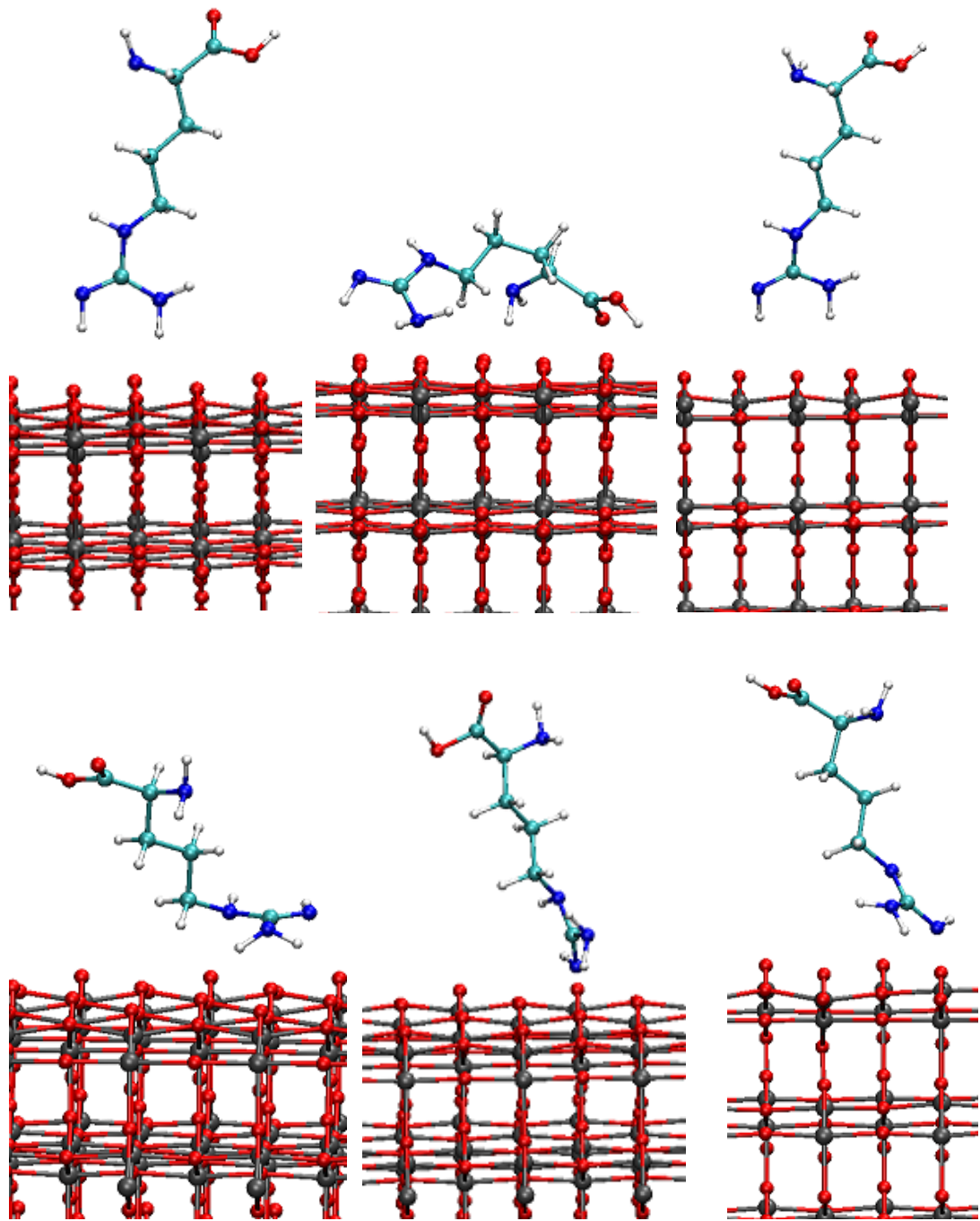

**Figure 6.** Optimized structure of Arg adsorbed on TiO$_2$ through the amine groups: configurations defined in the text as HO (top row) and NTi (bottom row) obtaibned with tiorg-0-1* (left), matsci-0-3 (middle) and with DFT (right).

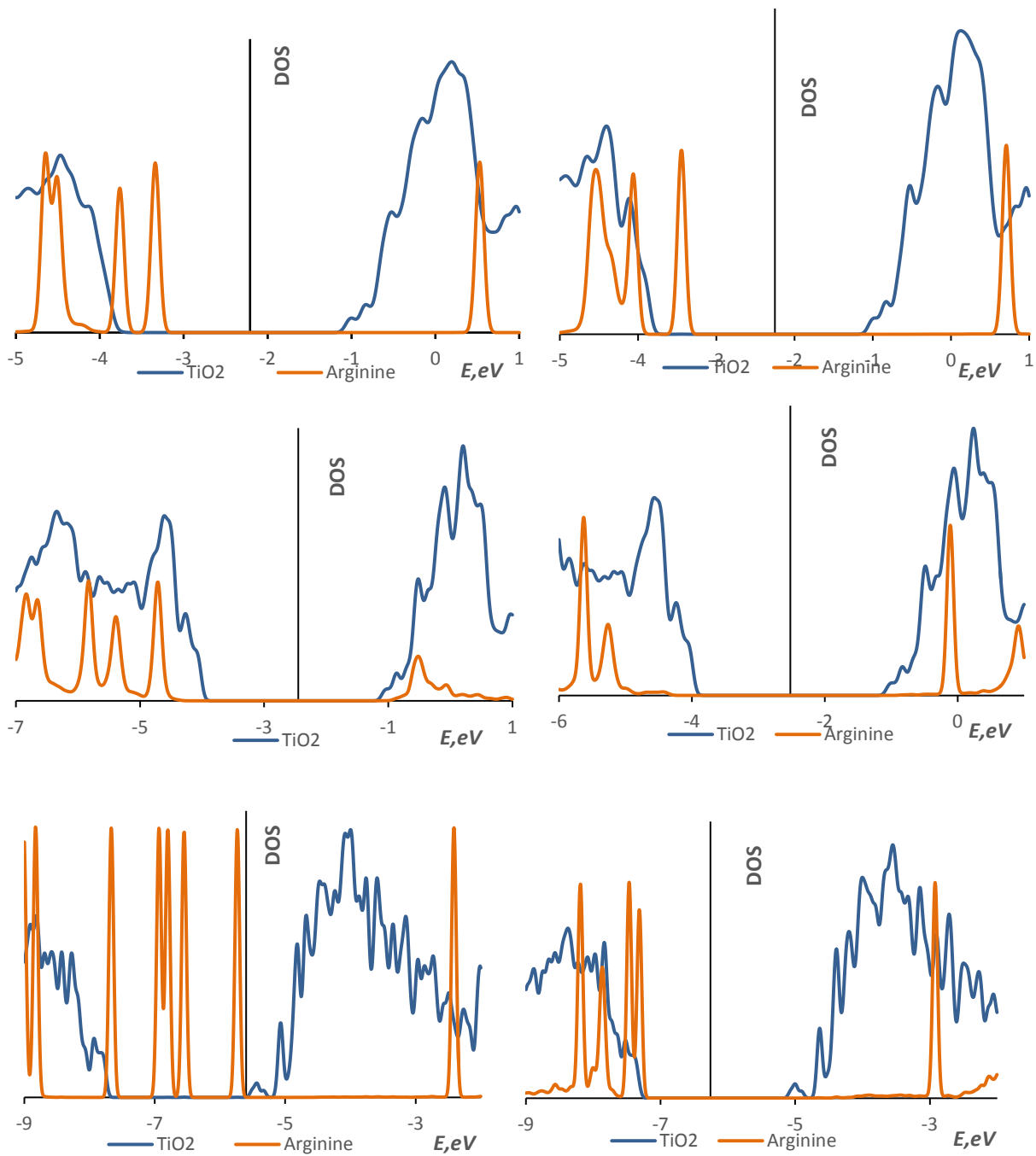

**Figure 7.** Molecule- and substrate- projected density of states of HO (left) and NTi (right) configurations of Arg on TiO₂ obtained in DFTB with the tiorg-0-1* parameters (top row) and matsci-0-3 DFT (middle row) and with DFT (bottom row). The *y* axis crosses the energy axis at the Fermi energy.

### 3.3. Arginine dipeptide -titania interfaces

We have computed the adsorption of Arginine dipeptide on titania with DFT and DFTB. Only adsorption via the COOH group was computed and for DFTB, only the tiorg-0-1* parameterization was used, as only at this juncture qualitative agreement was observed for Arginine adsorption considered above. This is sufficient for our purposes. The adsorption geometries obtained with both methods are shown in Figure 8; adsorption energies and key bond lengths are listed in Table 4. The level of agreement between DFT and DFTB is similar to that obtained for monomer adsorption considered above, in terms of both adsorption energies and bond lengths at the interface. The order of adsorption energies is somewhat violated in DFTB in that in predicts the weakest adsorption in M1 configuration, but the differences on the order of 0.1 eV are similar to what was seen with the monomer.

**Table 4.** Adsorption energies $E_{ads}$ (in eV) of Arginine dipeptide in different configurations on anatase (101) surface of $TiO_2$. The bond length for bonding between the molecule's and surface atoms are also given (in Å). For bidentate configurations, the two bond lengths are $O_{mol}$-Ti; for monodentate, they are $O_{mol}$-Ti and H-$O_{surf}$.

| System | $E_{ads}$, eV | $O_{mol}$-Ti, Å | $O_{mol}$-Ti/H-$O_{surf}$, Å |
|---|---|---|---|
| DFTB/Tiorg-0-1* | | | |
| BB | -1.01 | 2.10 | 2.11 |
| M1 | -0.92 | 1.99 | 1.00 |
| M2 | -1.13 | 1.99 | 0.99 |
| DFT/PBE | | | |
| BB | -0.88 | 2.07 | 2.05 |
| M1 | -1.13 | 2.13 | 1.54 |
| M2 | -1.07 | 2.18 | 1.47 |

Figure 9 shown the partial densities of states for all adsorption configurations obtained with the two methods. Similarly to the monomer case, there is a qualitative agreement in the band alignment in that the molecular HOMO enters the band gap (about 0.1 and 0.2 eV above the VBM for M1 and M2, respectively) in the monodentate configurations, while in the BB regime it remains slightly lower than the VBM (by 0.03 eV). We note that with DFT, the HOMO becomes very close to the conduction band minimum, lower than CBM by only 0.23, 0.25, and 0.24 eV in BB, M1, and M2 configurations, respectively.

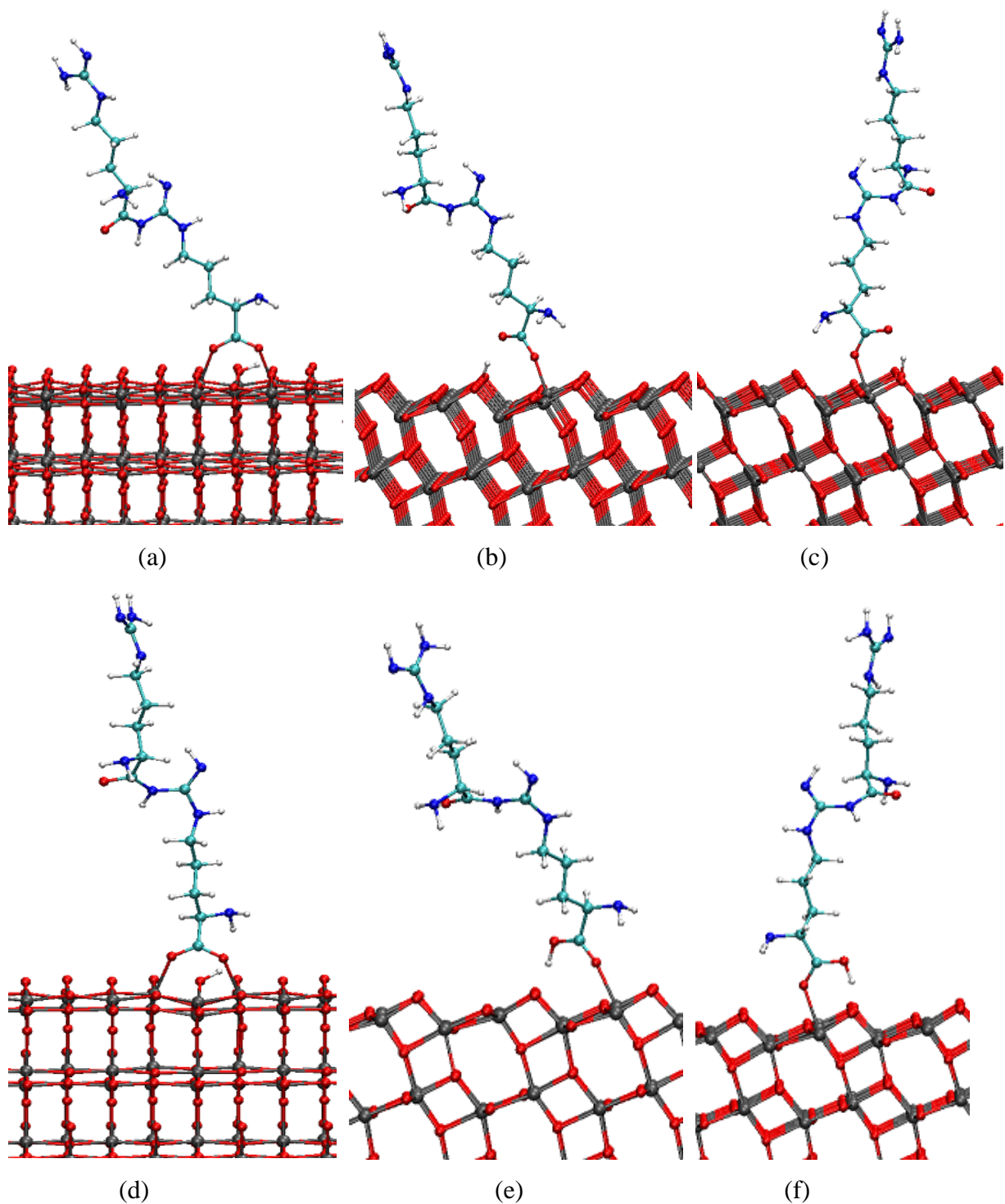

**Figure 8.** Optimized structures of BB (a), M1 (b) and M2 (c) for arginine dipeptide adsorbed on anatase TiO$_2$ obtained with DFTB tiorg-0-1* and BB (d), M1 (e) and M2 (f) from SIESTA.

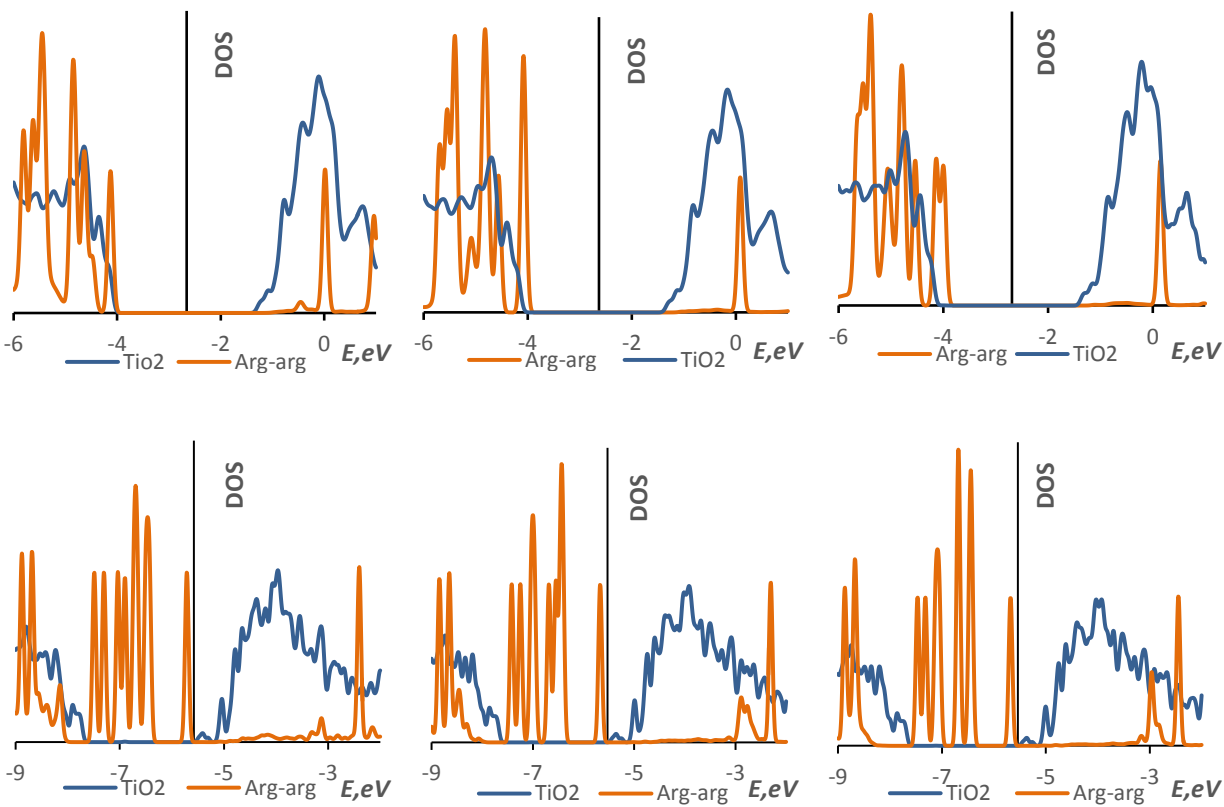

**Figure 9.** Molecule- and substrate- projected density of states of BB, M1 and M2 configurations of Arg$_2$ on TiO$_2$ obtained in DFTB with the tiorg-0-1* setup (top row) and with DFT (bottom row). The *y* axis crosses the energy axis at the Fermi energy.

### 3.4. TAT-titania interface

We have computed the adsorption of TAT on TiO$_2$ in DFTB (with the tiorg-0-1* parameters) via a –COOH group. The adsorption geometries are shown in Figure 10 and adsorption energies are listed in Table 5. Due to a large number of angular degrees of freedom in the peptide with shallow potential, the force convergence of 0.02 eV/Å was not achieved here; we have confirmed that the energy and the DOS were converged. Due to the large size of the molecule, other aminoacids contribute to binding to the surface, as can be seen in the figure, which contributes to a strengthening of the adsorption compared to Arg and Arg$_2$. The key O$_{mol}$-Ti bond lengths are similar to those obtained with the small molecules. What interests us here is the band alignment between the molecule and the surface. In Figure 11(a-c), we show the PDOS for these adsorbate systems. Similar to the case of individual peptide, the molecular HOMO is in the bandgap and LUMO in the conduction band. To compare this band alignment to that obtained with DFT/PBE, due to the prohibitive computational cost of the TAT/$TiO_2$ system (models shown in Figure 10 have more than 2500 atoms), we compare the DOS of the molecule and the anatase (101) surface computed individually and aligned based on the alignment obtained with Arg and $Arg_2$ and the known destabilization of TAT's HOMO vs the HOMO of Arg and $Arg_2$. This band alignment is shown in the last panel of Figure 11. As was expected from the analysis of Figures 3 and 5 together, the molecular HOMO enters the conduction band. Indeed, the HOMO of free TAT is 0.91 eV above that of Arg (with PBE, see Figure 3), and the HOMO of Arg on the anatase (101) surface is only about 0.6-0.7 eV lower than the CBM of the oxide (see Figure 5). The HOMO of free TAT is about 0.76 eV above that of $Arg_2$ (with PBE, see Figure 3), and the HOMO of $Arg_2$ on the anatase (101) surface is only about 0.24 eV lower than the CBM of the oxide (see Figure 9). It is therefore expected that with DFT/PBE, TAT's HOMO would enter the CB and this is exactly what is seen in Figure 11(d). This band alignment would have significant practically important consequences: the molecule would be oxidized.

**Table 5.** Adsorption energies $E_{ads}$ (in eV) of TAT in different configurations on anatase (101) surface of $TiO_2$. The bond length for bonding between the molecule's and surface atoms are also given (in Å). For bidentate configurations, the two bond lengths are $O_{mol}$-Ti; for monodentate, they are $O_{mol}$-Ti and H-$O_{surf.}$

| System | $E_{ads}$, eV | $O_{mol}$-Ti, Å | $O_{mol}$-Ti/H-$O_{surf}$, Å |
|---|---|---|---|
| DFTB/Tiorg-0-1* | | | |
| BB | -1.28 | 2.11 | 2.11 |
| M1 | -0.99 | 1.98 | 1.00 |
| M2 | -1.37 | 1.99 | 0.99 |

The electronic structure resulting from DFT/PBE is therefore qualitatively different from that resulting from DFTB. However, in this case it is not the DFTB but the DFT calculation which must be wrong. The high quality DFT calculations of TAT in section 3.1 (in Gaussian 09 and using the B3LYP functional) predict a HOMO level of about -5.0 eV, as can be seen in Figure 3.

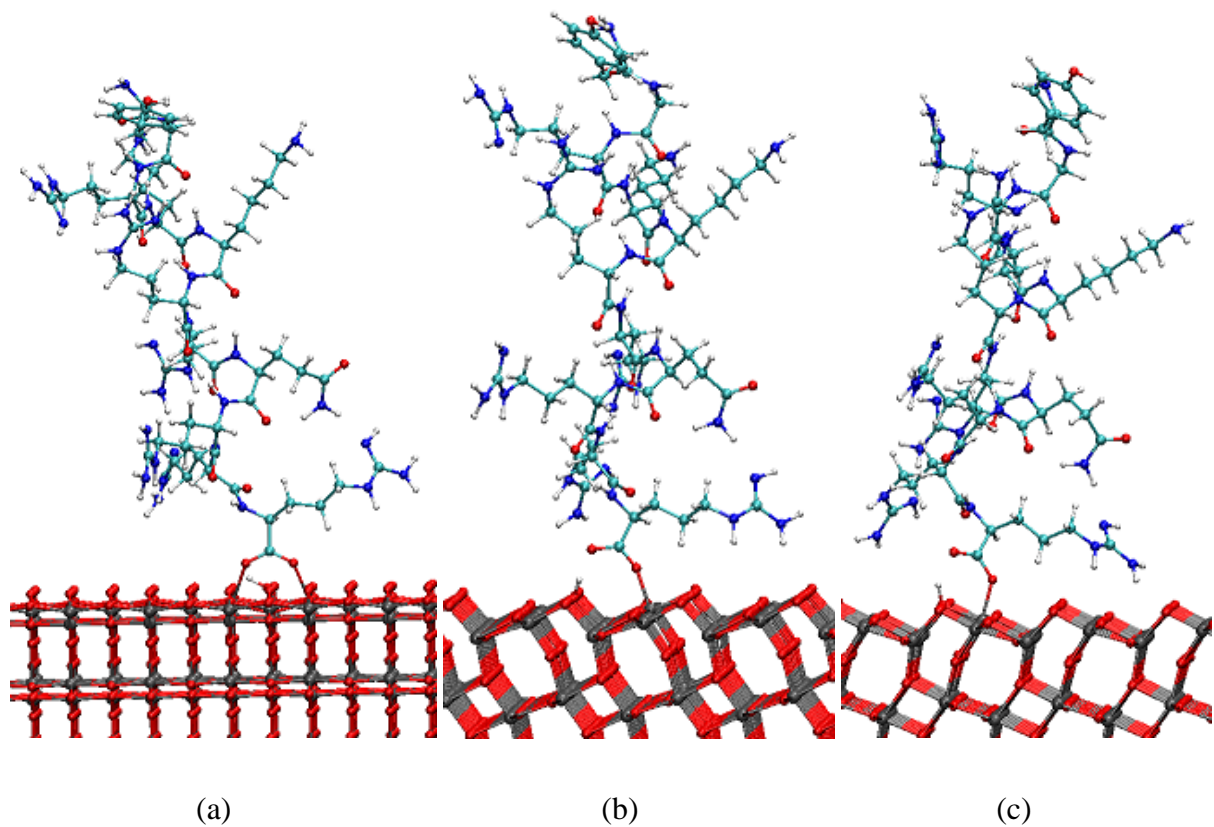

**Figure 10.** Optimized structures of BB (a), M1 (b) and M2 (c) configurations of TAT adsorbed on anatase TiO$_2$ obtained with DFTB using tiorg-0-1* parameter set.

While computational studies using hybrid functionals widely vary in their estimates of the CBM of anatase and its clean (101) surface[124-126], the latest high-quality experimental[126-127] and hybrid functional based DFT calculations using the HSE06[128-129] functional put the CBM of the anatase (101) surface at -5.10 eV.[126 130] We computed the HOMO of TAT with HSE06 (in Gaussian 09) and obtained a value of -5.20 eV. That is to say, the HOMO of TAT should remain below the CBM of anatase (101). The DFTB achieves that, in part because the DFTB parameterization was done in a way that effectively reproduced the bandgap of TiO$_2$.[86] The band alignment is expected to be correctly reproduced with an appropriate hybrid functional, but, as explained in the Introduction, such calculations are not practical for modeling of interfaces involving large biomolecules, due to the high CPU cost of computing exact exchange. The practical DFT approach for large interfaces remains that using the GGA approximation. In Ref. [84], we showed that DFTB with matsci-0-3 and tiorg-0-1* parameterizations does not produce a qualitatively correct band alignment of a *small* dye adsorbed on the anatase (101) surface of titania, while DFT/PBE does. It was not surprising to find a system where DFTB fails; after all, it is an approximation to DFT which is not expected to perform well in all cases.

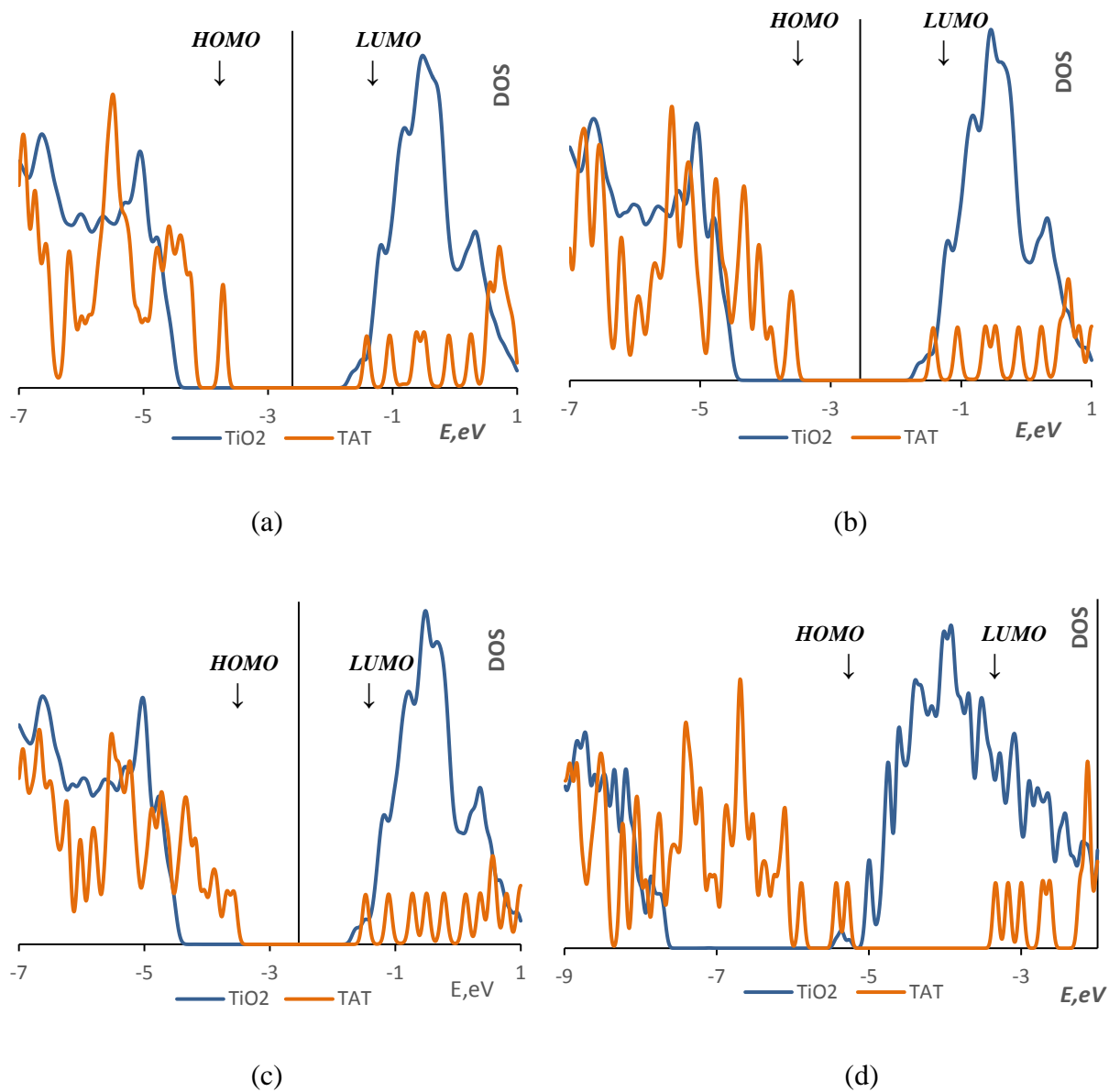

**Figure 11.** (a-c) Molecule- and substrate- projected density of states of BB, M1 and M2 configurations of TAT on $TiO_2$ obtained in DFTB with the tiorg-0-1* parameters. The *y* axis crosses the energy axis at the Fermi energy. (d) The simulated band alignment between the TAT and the (101) anatase surface. Approximate positions of molecular HOMO and LUMO energies are indicated with arrows.

Here, we have a somewhat less expected situation where DFT/PBE fails, but DFTB does not. *This is specific to the adsorption of large molecules (i.e. all biomolecules)* where HOMO is closer to the CBM. This makes the system more sensitive to the description of the band gap of the semiconductor substrate, which is known to be erroneous with GGA[125,131-134]. A practically important conclusion of the above is that DFTB is a good approach to model biomolecule-semiconductor interfaces, and that is not only due to the CPU cost advantage of three orders of magnitude but also because it can effectively, via parameterization, achieve qualitatively correct band alignment which may not be possible with GGA.

## 4. Conclusions

We have conducted a comparative computational study of interactions of biomolecules with an oxide surface. Such mixed bioinorganic systems present difficulties for ab initio modelling due to their sheer size. Specifically, the large size of biological molecules necessitates the use of large slabs to model the substrates, which makes more accurate ab initio schemes impractical. For example, the TAT/$TiO_2$ model used here has more than 2500 atoms. For most laboratories, the only practical and relatively accurate ab initio approach to model such system remains DFT with the GGA approximation. Even with a GGA functional (i.e. without the additional cost of exact exchange), the CPU cost is substantial. The DFTB method – an approximate density functional based method - therefore looks very promising for the modeling of such system, offering a three orders of magnitude speedup. Previous works reported both successes and failures of DFTB with specific parameterizations for the modeling of organic-inorganic interfaces.

In this work, we have compared the performance of DFT to DFTB for the modeling of adsorption on the widely use substrate $TiO_2$ of *both a small* biomolecule (an aminoacid, Arg) and a *large* biomolecule (the TAT cell-penetrating peptide). We have found that the quality of DFTB calculations depends on their ability to reproduce qualitatively the correct band alignment between the molecule and the surface, which in this case signifies the molecular HOMO in the titania bandgap and LUMO in the conduction band. Specifically, for systems and with a parameterization where a correct band alignment is obtained, we obtained a decent agreement in adsorption energies and geometries between DFT and DFTB.

Importantly, by comparing absorption of biomolecules of different size – an aminoacid, a dipeptide, and a real-sized peptide – we have discovered a seemingly counterintuitive phenomenon whereby it is the GGA DFT that fails while the DFTB is at least qualitatively correct. Specifically, due to a large destabilization of the HOMO of a large peptide vs a single aminoacid combined with an underestimation of the titania bandgap typical of GGA functionals, DFT is predicted to result in HOMO entering the conduction band, which would effectively ionize the molecule. DFTB, on the other hand, is able to reproduce a correct bandstructure precisely because it is an approximate method with which one can *effectively* reproduce the correct band gap. This effect is specific to large molecules such as biomolecules and may not have been appreciated due to a tendency to perform ab initio modeling on simplified, abridged systems. To get a correct band alignment for such a bioinorganic interface with DFT would require the use of (range-separated) hybrid functionals, which is prohibitive for most labs. This highlights the utility of the DFTB method for the modeling of bioinorganic interfaces not only from the CPU cost perspective but also from the accuracy point of view.

**Acknowledgements**

We thank Prof. Peter Deak from the University of Bremen for discussions about modeling the band structure of titania, Prof. Thomas Frauenheim and Dr. Balint Aradi from the University of Bremen for discussions about the DFTB method. Dr. Vadym Kulish and Dr. Mahasin Alam SK from NUS as well as Zhang Zhiwei, University of Illinois – Urbana Champagne are thanked for assistance with some calculations.